# First-principles study of high-temperature superconductivity in $X_2MH_6$ compounds under 20 GPa

Jing Luo[1], Qun Wei[1,*], Xiaofei Jia[1], Meiguang Zhang[2,*], Xuanmin Zhu[3]

[1] School of Physics, Xidian University, Xi'an 710071, China

[2] College of Physics and Optoelectronic Technology, Baoji University of Arts and Sciences, 721016 Baoji, China

[3] School of Information, Guizhou University of Finance and Economics, Guiyang 550025, China

E-mail: qunwei@xidian.edu.cn and zhmgbj@126.com.

**Abstract**

Research on high-temperature superconductors has primarily focused on hydrogen-rich compounds, however, the need for extreme pressures limits their practical applications. The $X_2MH_6$-type structure $Mg_2IrH_6$ stands out because it exhibits superconductivity at 160 K under ambient pressure. This study explores ways to enhance the superconducting critical temperature of this structure through atomic substitution and low-pressure treatment. It also evaluates the mechanical, thermodynamic, and dynamic stability of structures formed by replacing Mg and Ir in $Mg_2IrH_6$ with elements from the same groups using first-principles calculations. The findings identify eleven stable ternary compounds, four of which exhibit superconducting critical temperatures, with two compounds, $Mg_2RhH_6$ and $Mg_2IrH_6$, exceeding 77 K, classifying them as high-temperature superconductors. Their superconducting figures of merit (S values) are 1.83 and 2.20, respectively, indicating significant potential for practical applications. Theoretical analysis reveals that mid-frequency hydrogen phonons play a crucial role in enhancing superconducting properties through strong electron-phonon coupling interactions. The band structure study highlights the importance of van Hove singularities near the Fermi level. In addition, electron localization function and Fermi surface topology analyses reveal that the Fermi surface shape and density of states are crucial for increasing superconducting critical temperatures.

## 1. Introduction

Since mercury was discovered to exhibit zero electrical resistance at 4.2 K, researches into superconductivity in condensed matter physics has improved rapidly. By exploring the microscopic mechanisms of superconductivity, the Bardeen-Cooper-Schrieffer theory [1] was developed to explain superconducting behavior at low temperatures. However, the critical temperature of most currently synthesized superconductors remains below the temperature of liquid nitrogen (77 K), thus, research on high-temperature superconductors is essential. Based on advanced crystal structure search methods [2-6], many hydrogen-rich compounds [7-9] under high pressures have been predicted to exhibit high critical temperatures. Among them, $H_3S$ [10], $LaH_{10}$ [11, 12], and $YH_6$ [13] have been experimentally confirmed to have critical temperatures above 200 K under high pressures, whereas $Li_2MgH_{16}$ [14], $Li_2NaH_{17}$ [15], and $Li_2Na_3H_{23}$ [15] have been predicted to exhibit room-temperature superconductivity under pressures. Researches on hydrogen-rich compounds indicates that superconducting critical temperatures close to room temperature can be achieved through the conventional mechanism of electron-phonon interactions [16, 17]. However, superconductors with excellent superconducting properties typically require a minimum stable pressure of 150 GPa [17]. The critical temperature of superconductors at 0 GPa is generally low, with $LaH_2$ having a critical temperature of only 1.17 K [18]. This condition significantly limits the applications of these materials.

Notably, the $Mg_2IrH_6$ structure proposed by Dolui et al. exhibits a superconducting critical temperature (160 K) under ambient pressure [19]. Zhang et al. have found that structures with the unique $MH_6$ motif, such as $Mg_2RhH_6$ and $Li_2CuH_6$, exhibit high potential to become materials with elevated superconducting critical temperatures ($T_c$) under environmental pressure [20]. To further enhance the critical temperature of this structure, we performed atomic substitution and applied additional pressure based on the structure. Based on the ternary convex hull provided by Dolui [19], $Mg_2IrH_6$ is thermodynamically stable at 20 GPa; therefore, we applied a pressure of 20 GPa to this structure. Here, atomic substitution was used to replace the Mg and Ir atoms in $Mg_2IrH_6$ with elements from the second and ninth groups at 20 GPa. By evaluating the mechanical, thermodynamic, and dynamic stability of the resulting structures, we identified eleven stable crystal configurations. After excluding a non-metallic structure from these stable structures, we conducted in-depth analysis of the electrical and superconducting properties of the ten obtained metallic structures to provide insights into the design of new superconductors.

## 2. Computational Methods

The Vienna Ab initio Simulation Package (VASP) was used to perform relaxation and simulate the related properties of the 15 substituted structures [21]. The projector-augmented wave potentials [22] and Perdew–Burke–Ernzerhof exchange-correlation functional under generalized gradient approximation [23] were adopted. A cut-off energy of 550 eV was applied for plane wave expansion, and the Monkhorst–Pack $k$-grid [24] with a spacing of $2\pi \times 0.02$ Å$^{-1}$ was used to ensure adequate convergence (1 × 10$^{-5}$ eV/atom) of the total energy. In addition, the single-crystal elastic constants were determined from the strain–stress relationships derived by applying six finite deformations [25]. Dynamic stability was evaluated via the finite displacement approach, with phonon spectra computed using the PHONOPY package [26]. The energy convergence criterion is set to 10$^{-8}$ eV/atom, with a Gaussian smearing width of 0.05 eV. The Hessian matrix is calculated using Density Functional Perturbation Theory (DFPT) [27]. Electron–phonon coupling (EPC) was analyzed using Quantum ESPRESSO [28], with kinetic energy cut-offs set at 50 Ry, and ultrasoft pseudopotentials were applied. A 24 × 24 × 24 $k$-grid was used to determine the self-consistent electron density, and a 6 × 6 × 6 *q*-grid was employed to calculate the EPC constants. The superconducting critical temperatures were calculated using the Allen-Dynes modified McMillan equation [29].

## 3. Results and Discussion

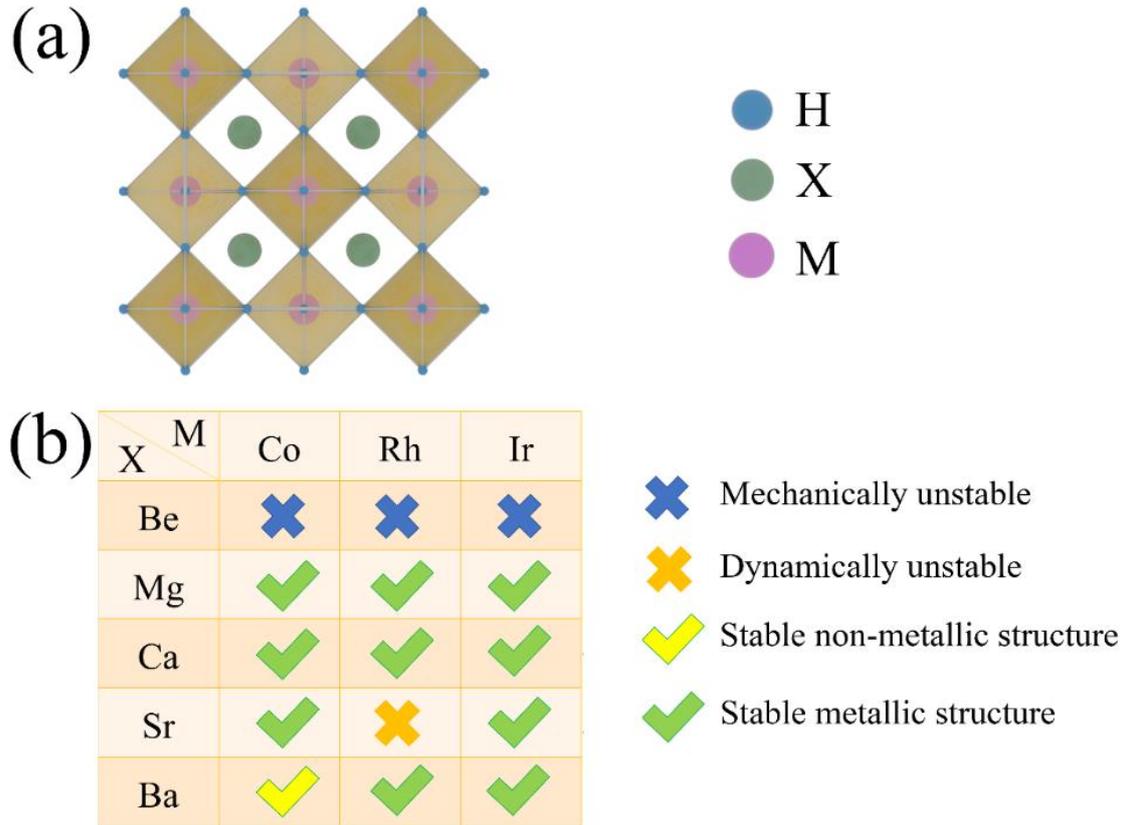

**Figure 1.** (a) Prototype structure of $X_2MH_6$. The X, M, and H atoms are coloured in green, red and blue, respectively. (b) Screening results.

The $X_2MH_6$ structure is a face-centred cubic phase characterized by the $MH_6$ octahedral structural unit, where X represents Be, Mg, Ca, Sr, and Ba atoms, and M represents Co, Rh, and Ir atoms (Figure 1). By substituting atomic positions, we obtained fifteen structures. Among them, three structures were found to be mechanically unstable, none were thermodynamically unstable, one structure exhibited dynamic instability, and one non-metallic structure was excluded. Consequently, ten stable metallic structures were identified. The specific process involves first calculating the elastic constants of these structures, followed by evaluating their mechanical stability according to the Born criteria [30], excluding those that do not meet the Born criteria. The elastic constants of the 15 structures were listed in Table S1 in the supplementary materials. For twelve mechanically stable structures, the formation energy $E_f$ is used to evaluate their thermodynamic stability, which can be calculated using the following formula [31, 32]:

$$E_f = [E(X_2MH_6) - 2E(X) - E(M) - 6E(H)]/9 \qquad (1)$$

where $E_f$ represents the formation energy of a material; $E(X_2MH_6)$ represents the total energy of the structure and $E(X)$, $E(M)$ and $E(H)$ represent the average energies of X, M, and H atom, respectively. The calculation results have been shown in Table S2 in the supplementary materials.

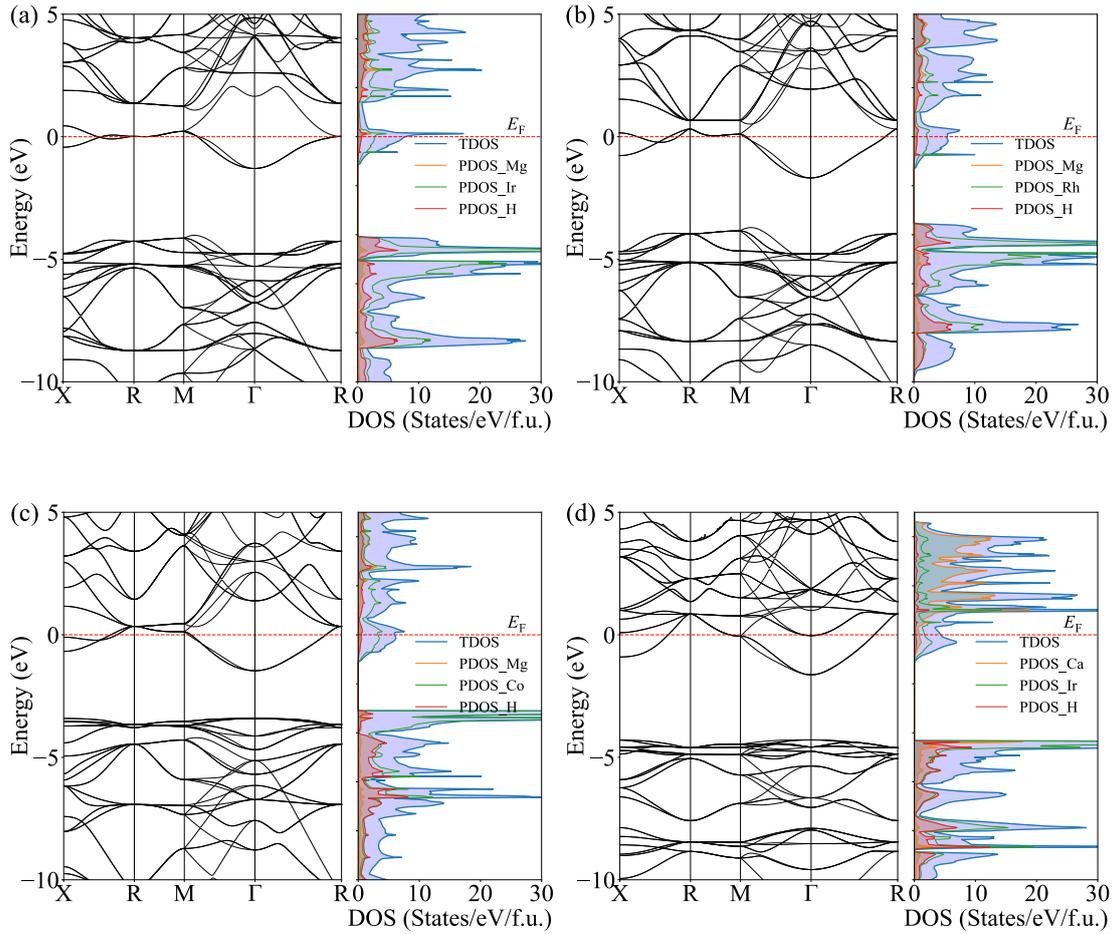

**Figure 2.** Calculated band structures and the DOS of the (a) $Mg_2IrH_6$, (b) $Mg_2RhH_6$, (c) $Mg_2CoH_6$, (d) $Ca_2IrH_6$ (The band structures and the DOS of the other six structures

are provided in Figure S1 in the supplementary file.)

Following the identification of 12 structures demonstrating mechanical stability and thermodynamic stability, we systematically evaluated their dynamical stability through comprehensive phonon spectrum calculations, specifically examining the potential existence of imaginary frequencies in the phonon spectra. This process ultimately identified 11 stable structures. Finally, 10 stable metallic structures were identified by calculating their band structures, as shown in Figures 2 and S1. The detailed screening results are presented in Figure 1(b). The ten stable metallic structures were as follows: $Mg_2CoH_6$, $Mg_2RhH_6$, $Mg_2IrH_6$, $Ca_2CoH_6$, $Ca_2RhH_6$, $Ca_2IrH_6$, $Sr_2CoH_6$, $Sr_2IrH_6$, $Ba_2RhH_6$, and $Ba_2IrH_6$.

For comparison, we also performed calculations on the mechanical and thermodynamic stabilities of two well-known high-temperature superconductors, $YCaH_{12}$ [33] and $ScYH_6$ [34]. The elastic constants and formation energies of these two structures were listed in Tables S1 and S2. The computational results demonstrate that both structures exhibit negative formation energies, while their satisfaction of the Born-Huang stability criteria confirms simultaneous thermodynamic and mechanical stability.

To further investigate the thermodynamic stability of these ten stable ternary metal hydrides $X_2MH_6$, we systematically investigated potential decomposition pathways by identifying related stable phases in the Open Quantum Materials Database (OQMD) [35, 36] and subsequently constructing the corresponding phase diagrams for $X_2MH_6$ structures. The thermodynamic stability of these structures was quantitatively evaluated by calculating their decomposition energy ($E_d$), defined as the enthalpy difference between the $X_2MH_6$ compound and its most stable decomposition products. A negative decomposition enthalpy ($E_d < 0$) signifies high thermodynamic stability, indicating that the compound is energetically favorable and can be synthesized through established reaction pathways. A positive decomposition enthalpy ($E_d > 0$) corresponds to reduced thermodynamic stability, suggesting a thermodynamic tendency for spontaneous decomposition. The calculated decomposition pathways and corresponding enthalpy values are systematically presented in Table 1. All the structures exhibit negative decomposition enthalpies ($E_d < 0$), confirming their thermodynamic stability and suggesting potential synthetic routes through precursor compounds.

**TABLE 1.** The decomposition pathways, decomposition energies $E_d$ (eV).

| structures | decomposition path | $E_d$ |
|---|---|---|
| $Mg_2IrH_6$ | $Mg_2IrH_6 = Mg_2IrH_5 + \frac{1}{2}H_2$ | -0.041 |
| $Mg_2RhH_6$ | $Mg_2RhH_6 = MgH_2 + MgRhH_2 + H_2$ | -1.434 |
| $Mg_2CoH_6$ | $Mg_2CoH_6 = Mg_2CoH_5 + \frac{1}{2}H_2$ | -0.537 |
| $Ca_2IrH_6$ | $Ca_2IrH_6 = Ca_2IrH_5 + \frac{1}{2}H_2$ | -1.042 |
| $Ca_2RhH_6$ | $Ca_2RhH_6 = \frac{1}{3}Ca_4Rh_3H_{12} + \frac{2}{3}CaH_2 + \frac{1}{3}H_2$ | -0.596 |
| $Ca_2CoH_6$ | $Ca_2CoH_6 = \frac{1}{3}Ca_4Co_3H_{12} + \frac{2}{3}CaH_2 + \frac{1}{3}H_2$ | -0.906 |
| $Sr_2IrH_6$ | $Sr_2IrH_6 = \frac{1}{3}Sr_2IrH_5 + \frac{1}{3}Sr_3Ir_2H_{12} + \frac{1}{3}SrH$ | -0.679 |
| $Sr_2CoH_6$ | $Sr_2CoH_6 = SrCoH_4 + SrH_2$ | -0.860 |
| $Ba_2RhH_6$ | $Ba_2RhH_6 = Ba_2RhH_5 + \frac{1}{2}H_2$ | -1.397 |
| $Ba_2IrH_6$ | $Ba_2IrH_6 = \frac{1}{3}Ba_2IrH_5 + \frac{1}{3}Ba_3Ir_2H_{12} + \frac{1}{3}BaH$ | -0.774 |

After identifying the stable structures, we used Quantum ESPRESSO to calculate the superconducting critical temperatures of the ten structures. Among these ten structures, only $Mg_2IrH_6$, $Mg_2RhH_6$, $Mg_2CoH_6$, and $Ca_2IrH_6$ are superconductors. The detailed data for these four superconductors are presented in Table 2. The critical temperatures of the structures are expressed using the Allen-Dynes modified McMillan equation [29]:

$$T_c = \frac{\omega_{log}}{1.20} \exp\left[\frac{-1.04(1+\lambda)}{\lambda(1-0.62\mu^*) - \mu^*}\right] \quad (2)$$

where $\omega_{log}$ denotes the logarithmic average phonon frequency, $\lambda$ denotes the EPC constant and $\mu^*$ denotes the Coulomb potential parameter. $\lambda$ can be determined by integrating the Eliashberg spectral function $\alpha^2 F(\omega)$ in the frequency space [37]:

$$\lambda = 2 \int \frac{\alpha^2 F(\omega)}{\omega} d\omega \tag{3}$$

Figure 3 shows the Eliashberg phonon spectral function $\alpha^2 F(\omega)$ and $\lambda$. $\mu^*$ was set to be 0.1. The calculated $\lambda$ and $\omega_{log}$, along with the calculated $T_c$ values, were presented in Table 2. The critical temperatures of Mg$_2$IrH$_6$ and Mg$_2$RhH$_6$ at 0 GPa [38] are also presented in Table 2. Table 2 also presents the superconducting critical temperatures of Mg$_2$IrH$_6$ proposed by Dolui et al. [19] as well as the critical temperatures of Mg$_2$IrH$_6$ and Mg$_2$RhH$_6$ at 0 GPa calculated by Sanna et al. [38] and Cerqueira et al. [39]. Based on the pressure-dependent study of mercury-based superconductors by Jha et al. [40], the observed enhancement of $T_c$ under applied pressure can be attributed to pressure-induced modifications in the crystal lattice vibrational modes, consequently promoting the elevation of superconducting transition temperature. As evidenced by the data presented in Table 2, the applied pressure demonstrates an enhancement in critical temperature relative to the previously reported values by Sanna et al. [38] and Cerqueira et al [39].

**TABLE 2.** EPC constant $\lambda$, logarithmic average phonon frequency $\omega_{log}$ (in K), density of states at the Fermi level $N_F$ (in eV$^{-1}$), superconducting figure of merit $S$ and superconducting critical temperature $T_c$ (in K) of the structures under pressures (in GPa).

| Structures | Pressure | $\lambda$ | $\omega_{log}$ | $N_F$ | $T_c$ | $S$ | Refs. |
|---|---|---|---|---|---|---|---|
| Mg$_2$IrH$_6$ | 20 | 1.33 | 955.77 | 14.59 | 96.3 | 2.20 | This work |
| Mg$_2$IrH$_6$ | 0 | | | | 77.0 | 1.94 | [38] |
| Mg$_2$IrH$_6$ | 0 | | | | 59.4 | | [39] |
| Mg$_2$IrH$_6$ | 0 | | | | 160.0 | | [19] |
| Mg$_2$RhH$_6$ | 20 | 1.05 | 1069.63 | 11.16 | 80.2 | 1.83 | This work |
| Mg$_2$RhH$_6$ | 0 | | | | 48.5 | 1.24 | [38] |
| Mg$_2$RhH$_6$ | 0 | | | | 53.8 | | [39] |
| Mg$_2$CoH$_6$ | 20 | 0.75 | 1111.20 | 10.35 | 45.8 | 1.044 | This work |
| Ca$_2$IrH$_6$ | 20 | 0.35 | 1200.29 | 9.87 | 2.2 | 0.05 | This work |

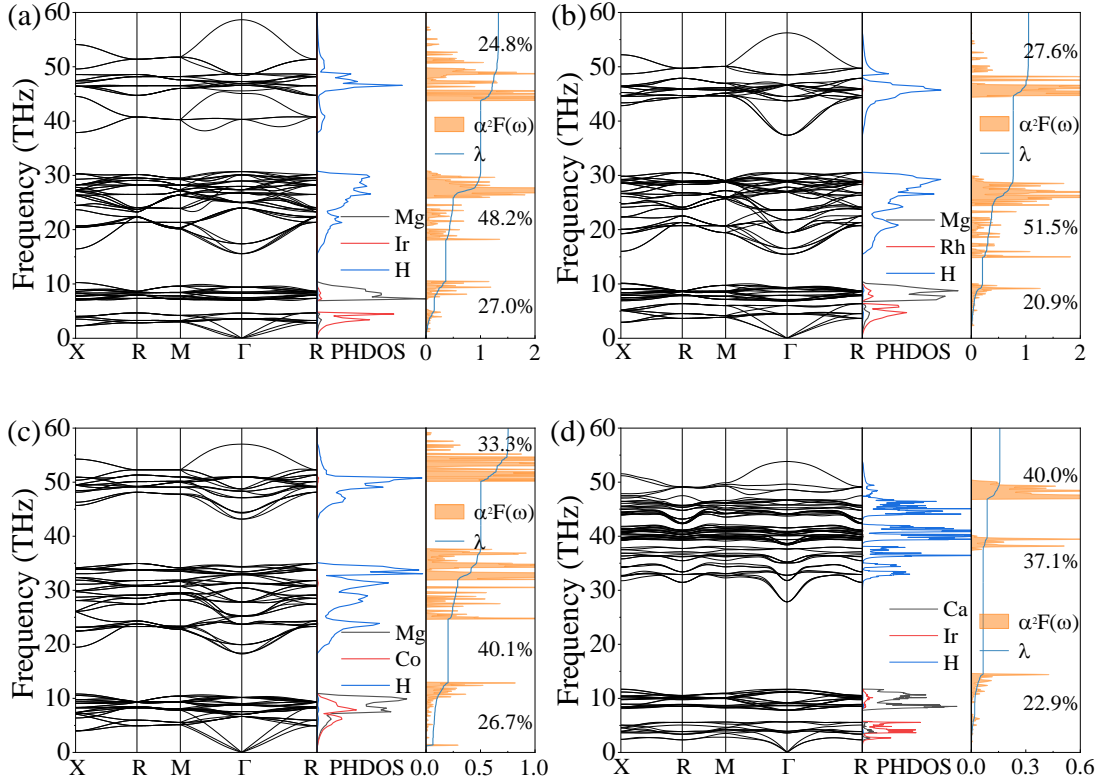

**Figure 3.** Calculated phonon dispersion curves, phonon density of states, Eliashberg phonon spectral function $\alpha^2F(\omega)$, EPC parameter $\lambda$, and the contribution of the phonon modes to the EPC for different frequency ranges of the (a) $Mg_2IrH_6$, (b) $Mg_2RhH_6$, (c) $Mg_2CoH_6$, and (d) $Ca_2IrH_6$. (The phonon spectra of the other six structures are provided in Figure S2 in the supplementary file.)

Among the four superconductors, the superconducting critical temperatures of $Mg_2IrH_6$ and $Mg_2RhH_6$ exceed 77 K, thereby classifying them as high-temperature superconductors. The superconducting figure of merit $S$ is used to comprehensively evaluate the performance and practicality of superconducting materials [41]. This metric not only considers the superconducting critical temperature $T_c$ but also incorporates the pressure conditions, reflecting the feasibility of superconducting materials in real-world applications. The superconducting figures of merit for all structures are presented in Table 2. $MgB_2$, as a material with technological applications, has a superconducting figure of merit of $S = 1$. The calculated $S$ values are 2.20 for $Mg_2IrH_6$ and 1.83 for $Mg_2RhH_6$, which are higher than that of $MgB_2$, indicating the greater potential for practical applications.

The critical temperatures of $Mg_2IrH_6$ and $Mg_2RhH_6$ under 20 GPa exceed 77 K. Therefore, studying the thermal stability of these materials at higher temperature is important. Thus, we carried out a 10 *ps ab initio* molecular dynamics (AIMD) simulation at 200 K and illustrated the results in Figure S3. The results of the calculation show that the energy of these structures remains relatively stable after 2 *ps*, indicating that these materials exhibit remarkable stability at 200 K, with no significant structural transition occurring.

EPC λ is a key factor influencing the superconducting critical temperature of superconductors. The EPC values of the four structures are primarily contributed by phonons in three frequency ranges: low-frequency range (<15 THz), mid-frequency range (15–40 THz), and high-frequency range (>40 THz). The contributions of each frequency range to the EPC are labeled in Figures 3. For example, for the high-temperature superconductor $Mg_2RhH_6$, the contributions of low-, mid-, and high-frequency ranges to the EPC are 20.9%, 51.5%, and 27.6%, respectively, as shown in Figure 3(b). For $Mg_2IrH_6$ and $Mg_2RhH_6$, phonons in the mid-frequency range contribute more significantly to the EPC constant, whereas the influence of high-frequency and low-frequency phonons on EPC is roughly comparable. In contrast, for $Mg_2CoH_6$ and $Ca_2IrH_6$, the contribution of mid-frequency phonons is relatively small, around 40%.

To further investigate the relationship between atoms and phonons, the phonon dispersion curves and phonon density of states (DOS) of the obtained structures are presented in Figure 3. The figures show that the low-frequency phonons are primarily contributed by the two metallic elements, whereas the mid- and high-frequency phonons are primarily contributed by hydrogen atoms. Therefore, for $Mg_2IrH_6$ and $Mg_2RhH_6$, mid-frequency hydrogen atoms predominantly contribute to EPC, thereby increasing the superconducting critical temperature. In contrast, for $Mg_2CoH_6$ and $Ca_2IrH_6$, the contributions from metal atoms and high-frequency hydrogen atoms to the EPC are more significant, while the absence of mid-frequency phonons may limit the enhancement of the superconducting critical temperature.

According to the study by Wang et al. [42], the vibrations of the $IrH_6^{4-}$ anion in the $X_2MH_6$ family are crucial for the superconducting mechanism. These vibrations induce electron coupling in the $e_g^*$ orbitals. The lower-energy $d$ orbitals of Ca in $Ca_2IrH_6$, compared to $Mg_2IrH_6$, lead to reversed $d$-orbital back-donation, suppressing this coupling mechanism and ultimately resulting in a lower superconducting critical temperature. This is consistent with our findings. Therefore, we believe that the $d$ orbitals of the X atom can interact with the $e_g^*$ orbitals of the $MH_6^{4-}$ anion, thereby altering the phonon energy of hydrogen and, in turn, affecting the critical temperature.

The electronic structure of materials is closely related to their superconductivity. Thus, the band structures and DOS of the obtained structures are showed in Figures 2 and S3. The electronic structure of the high-temperature superconductors exhibits distinctive features near the Fermi level. For example, the $Mg_2IrH_6$ structure exhibits a van Hove singularity near the Fermi level, characterized by an energy peak, which could explain its highest superconducting critical temperature. In contrast, the lower critical temperatures of $Mg_2RhH_6$ may be due to the presence of van Hove singularities near the Fermi level but with energy valleys instead of peaks. Compared to $Mg_2IrH_6$ and $Mg_2RhH_6$, the van Hove singularity in $Mg_2CoH_6$ and $Ca_2IrH_6$ and other non-superconducting $X_2MH_6$ in this study is farther from the Fermi level; thus, it does not significantly increase the DOS at the Fermi level. This may explain the lower superconducting critical temperatures of $Mg_2CoH_6$ and $Ca_2IrH_6$.

Additionally, the difference between high-temperature superconductors ($Mg_2IrH_6$ and $Mg_2RhH_6$) and low-temperature superconductors ($Mg_2CoH_6$ and $Ca_2IrH_6$) or non-superconducting metals lies in the contribution of hydrogen atoms at the Fermi level. The hydrogen content at the Fermi level is a significant factor in increasing the superconducting critical temperature [43]. According to the density of state shown in Figure 2, in all structures, the Fermi level is primarily dominated by M atoms. However, for $Mg_2IrH_6$ and $Mg_2RhH_6$, the contribution of H atoms is greater than or equal to that of X atoms, whereas, in the $Mg_2CoH_6$, $Ca_2IrH_6$ and non-superconducting materials in this study, the contribution of H atoms is lower and significantly less than that of X atoms. According to the theory of Wang et al. [42], the reason for this phenomenon may be related to the electronic orbitals of Mg and other atoms. The core states of Mg ($2s^2$, $2p^6$) are low in energy, while the empty states ($3s$, $3p$) are high. As a result, the frontier molecular orbitals of the $MH_6^{3-}$ anion in the solid are expected to be continuous, leading to a higher DOS near the Fermi level in the $Mg_2MH_6$ structures. In contrast, Ca, Sr, and Ba atoms have lower-lying $d$-orbitals, which deplete the charge in $MH_6^{3-}$, effectively reducing the superconducting critical temperature.

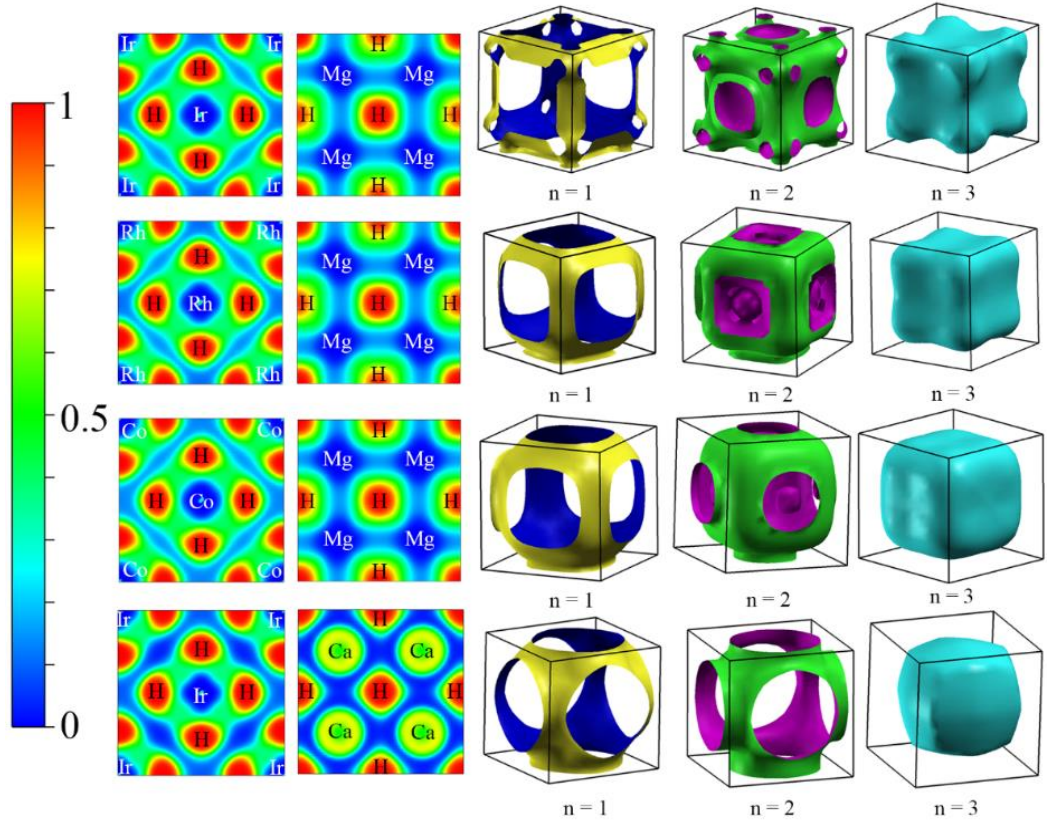

**Figure 4.** Electronic localization function (ELF) and Fermi surface topology of superconductors. (The ELF and Fermi surface topology of the other six structures are provided in Figure S4 in the supplementary file.)

To explore additional factors affecting the critical temperature, we plotted the Electron Localization Function (ELF) and Fermi surface topologies of these structures. Figure 4 and S4 present the ELF and Fermi surfaces of the $X_2MH_6$. The ELF plots of these structures show a minimal overlap of electron clouds between metal atoms and between metal and H atoms, indicating the absence of strong interactions between them. In contrast, the ELF values between H atoms are approximately 0.5, suggesting that some H atoms form weak covalent bonds. According to our calculations, the distance between hydrogen atoms is 2.418 Å. Based on the structure diagram of $X_2MH_6$, it can be determined that the H atoms form an octahedral hydrogen cage with a length of 2.418 Å, enclosing the M atom.

Comparing the ELF between the X and H atoms, the larger red region surrounding the H atom in the $Mg_2IrH_6$ and $Mg_2RhH_6$ indicates stronger electron localization, which promotes the formation of Cooper pairs, thereby enhancing EPC and increasing the superconducting critical temperature. In contrast, the green region surrounding the X atom in the $Mg_2CoH_6$, $Ca_2IrH_6$ and non-superconducting materials in this study suggests an increase in electron delocalization, resulting in a reduction in the EPC strength.

The superconducting properties of these structures are closely related to the topology of the Fermi surface. All $X_2MH_6$ structures have three conduction bands at the Fermi surface, with the first band labeled as n = 1. At n=1, the Fermi surfaces of all structures exhibit an open centre and closed edges, indicating that there may be numerous open orbits in the Brillouin zone, with minimal restriction on electron movement, leading to improved electronic conductivity. At n = 3, the Fermi surfaces of all structures display a closed and smoother shape, indicating that the structures exhibit typical metallic characteristics. However, at n = 2, significant differences in Fermi surface topology are observed between high-temperature superconductors ($Mg_2IrH_6$ and $Mg_2RhH_6$) and other structures in this study. The Fermi surface of $Mg_2CoH_6$, $Ca_2IrH_6$ and non-superconducting materials in this study at n = 2 is similar to that at n = 1, whereas the Fermi surface of the $Mg_2IrH_6$ and $Mg_2RhH_6$ is significantly more complex, featuring multi-layered structures and closed shapes. This suggests that the electron distribution is more intricate, with stronger orbital overlap, resulting in an increased DOS near the Fermi level. In general, a higher DOS near the Fermi surface favours the formation of Cooper pairs, which enhances EPC and increases the superconducting critical temperature.

**4. Conclusion**

In summary, to enhance the superconducting transition temperature of high-temperature superconductor $Mg_2IrH_6$, we combined atomic substitution and applied a pressure of 20 GPa, yielding 15 ternary structures. By substituting metal atoms from Groups II and IX and screening based on formation energy, the Born criterion, and phonon stability, we identified 11 stable structures, 10 of which exhibit metallic properties. At 20 GPa, the superconducting critical temperatures of the high-temperature superconductors $Mg_2IrH_6$ and $Mg_2RhH_6$ reached 96.3 K and 80.2 K, respectively, whereas the critical temperatures of the $Mg_2CoH_6$ and $Ca_2IrH_6$ were below 77 K. We observed that mid-frequency phonons (15-40 THz), particularly those associated with hydrogen, significantly affect EPC, thereby increasing the critical temperature. In addition, the band structure and ELF analyses revealed that the proximity of the van Hove singularity to the Fermi level and the contribution of hydrogen atoms at the Fermi level are key factors influencing superconducting performance. Furthermore, the complex multilayered topology of the Fermi surface increases the DOS, further increasing the superconducting critical temperature.

**Data availability statement**

All data that support the findings of this study are included within the article (and any supplementary files).

**Acknowledgment**

This work was financially supported by the National Natural Science Foundation of China (Grant Nos.: 11965005 and 11964026), the Natural Science Basic Research plan in Shaanxi Province of China (Grant Nos.: 2023-JC-YB-021, 2022JM-035). All the authors thank the computing facilities at High Performance Computing Center of Xidian University.